\documentclass[twocolumn,twocolappendix,iop,numberedappendix,appendixfloats]{openjournal}

\usepackage[utf8]{inputenc}

\usepackage[british]{babel}

\usepackage{amsmath}
\usepackage{amssymb}
\usepackage{amsfonts}
\usepackage{graphicx}
\usepackage{bm}
\usepackage{natbib}
\usepackage[colorlinks,allcolors=blue]{hyperref}

\usepackage{subfigure}
\usepackage{xspace}
\usepackage{multirow}
\usepackage{caption}
\usepackage{subcaption}
\usepackage{fontawesome}
\usepackage{placeins}

\usepackage{orcidlink}

\begin{document}

\journalinfo{The Open Journal of Astrophysics}

\shorttitle{SKA-Low simulations for a CD/EoR deep field}
\shortauthors{A. Bonaldi et al.}

\title{SKA-Low simulations for a cosmic dawn/epoch of reionisation deep field}


\author{Anna Bonaldi$^{1, \star}$}
\author{Philippa Hartley$^{1}$}

\author{Simon Purser$^{1}$}
\author{Omkar Bait$^{1}$}

\author{Eunseong Lee$^{2,3}$}
\author{Robert Braun$^{1}$}
\author{Florent Mertens$^{4,5}$}
\author{Andrea Bracco$^{6,7}$}
\author{Wendy Williams$^{1}$}
\author{Cath Trott$^{8,9}$}

\affiliation{$^{1}$SKA Observatory, Jodrell Bank, Lower Whitington, Macclesfield, SK11 9FT, UK}
\affiliation{$^{2}$Jodrell Bank Centre for Astrophysics, School of Physics and Astronomy, University of Manchester, Oxford Road, Manchester M13 9PL, U.K}
\affiliation{$^{3}$Department of Physics and Astronomy, University of Pennsylvania, Philadelphia, PA 19104, U.S.A.}
\affiliation{$^{4}$LUX, Observatoire de Paris, PSL Research University, CNRS, Sorbonne Universit\'e, F-75014 Paris, France}
\affiliation{$^{5}$Kapteyn Astronomical Institute, University of Groningen, PO Box 800, 9700 AV Groningen, The Netherlands}
\affiliation{$^{6}$INAF – Osservatorio Astrofisico di Arcetri, Largo E. Fermi 5, 50125 Firenze, Italy}
\affiliation{$^{7}$Laboratoire de Physique de l'Ecole Normale Sup\'erieure, ENS, Universit\'e PSL, CNRS, Sorbonne Universit\'e, Universit\'e de Paris, F-75005 Paris, France}
\affiliation{$^{8}$International Centre for Radio Astronomy Research, Curtin University,  Bentley WA, Australia}
\affiliation{$^{9}$ ARC Centre of Excellence for All-Sky Astrophysics in 3D, Australia}

\thanks{$^\star$ E-mail: \nolinkurl{anna.bonaldi@skao.int}}

\begin{abstract}
We present a realistic simulation of an SKA-Low cosmic dawn/epoch of reionisation (CD/EoR) observation, which can be used to further the development of foreground-mitigation approaches. The simulation corresponds to a deep (1000\,h) integration of a single pointing over the 106\,MHz--196\,MHz frequency range. The sky components include the CD/EoR signal, extragalactic foreground emission featuring strong (over 5\,Jy at 150\,MHz) out-of-field sources and in-field sources down to 1\,$\mu$Jy at 150\,MHz, and Galactic emission from the GSM2016 model complemented with small-scales structure beyond its native $\sim 1$\,deg resolution from a magneto-hydrodynamic simulation of the interstellar medium. Modeled errors include a partial de-mixing of the out-of-field sources, direction-dependent calibration errors leading to residual ionospheric effects, and direction-independent gain calibration errors, on top of thermal noise. Simulated observations are delivered as visibilities as well as imaging products with natural weighting. The true, uncorrupted, CD/EoR signal is also delivered, to allow an assessment of the efficacy of foreground-mitigation approaches. The codes used to generate these simulations are also delivered, so that new simulated datasets can be produced. This simulation has been the basis for the SKA Science Data Challenge 3a (SDC3a), which addressed foreground removal. 
\end{abstract}

\keywords{astro-ph.CO (Physics - Cosmology and Extragalactic Astrophysics), astro-ph.IM (Physics - Instrumentation and Methods for Astrophysics)}

\maketitle
\section{Introduction}
The detection of the 21 cm line from neutral hydrogen (HI) during the cosmic dawn (CD) and the epoch of reionisation (EoR) is a high priority science objective for the Square Kilometre Array Observatory \citep[SKAO,][]{koopmans15} and is being actively pursued by total power (EDGES \citealt{bowman08, bowman18}; SARAS \citealt{singh17, nambissan21}) and interferometric (GMRT \citealt{paciga11}; PAPER \citealt{kolopanis19}; LOFAR \citealt{patil17, mertens20}; MWA \citealt{barry19, li19, trott20, ewallwice16, yoshiura21}; HERA \citealt{abdurashidova22}; OVRO-LWA \citealt{eastwood19, garsden21}; NenuFAR; \citealt{munshi24})
experiments. 

During the CD, one of the least explored epochs of the Universe, starting at $z \sim 30$ ($\nu \sim 45$\,MHz) the neutral hydrogen provides a unique probe of the emergence of the first luminous sources and their impact on the surrounding medium. Subsequently, during the epoch of reionisation (EoR) tracking the abundance of HI at $\nu \sim $100--250\,MHz ($z \sim$13--4.7) gives us a detailed account on when and how the Universe globally transitioned from neutral to ionised. 

So far, the detection has been elusive, although great progress has been made on upper limits to the signal \citep[][and references therein]{2019arXiv190912491T}. One of the biggest challenges is the presence of radio emission from sources along the line of sight, due to external galaxies as well as our Milky Way, three to four orders of magnitude brighter than the signal of interest. This foreground emission needs to be strongly mitigated (with either subtraction or avoidance methods) before any measurement of the CD/EoR signal can be made \citep[see][for a review]{chapman19}. 

Foreground emission, dominated by synchrotron and free-free processes, exhibits an intrinsically smooth frequency behavior, whereas the 21-cm signal from the EoR typically decorrelates over a few MHz. This fundamental difference underpins many of the foreground removal methods developed in recent years \citep[e.g.][and references therein]{chapman2015,Mertens24}.

These methods have been shown to perform well in simulations. However, the real challenge arises when instrumental and environmental effects are taken into account. The chromaticity of the instrument, both in the point spread function and in the primary beam \citep{Trott2016,Ewall-Wice2016,Wilensky2024,Chokshi2024} separates smooth-spectrum foregrounds from frequency-dependent structures, making them much harder to isolate and subtract. Additional complications such as ionospheric fluctuations \citep{2018ApJ...867...15T,Brackenhoff2024,PalDattaMazumder2024,2026A&C....5401012H}, Radio-Frequency Interference \citep[RFI, e.g.][]{2023ApJ...957...78W}, and imperfect calibration can further distort the spectral behavior of foregrounds, introducing features that mimic or obscure the EoR signal. Achieving the accuracy needed for reliable detection therefore requires a joint treatment of foreground characterization and instrumental calibration \citep[e.g.][]{2016MNRAS.461.3135B}.

In this work we aim to bring simulation work close to the realistic scenario, by providing a detailed simulation of an SKA-Low EoR observation. Our simulation includes both diffuse and point-like foreground contamination, as well as the residual effect of out-of-field strong sources that enter the field of view (FoV) due to the telecope's sidelobes. We further model ionospheric effects and calibration errors. 

This is the most sophisticated SKA-Low EoR simulation produced to date, and we make it available, together with the codes used to produce it, as a benchmark against which foreground mitigation approaches can be tested and refined. This simulation has been the basis for the SKA science data challenge 3a \citep[SDC3a,][]{sdc3a2}, an EoR foreground mitigation exercise undertaken by the SKA science community in preparation for the advent of the SKA data.  

This paper is organized as follows. In Sec. \ref{sec:instrument} we describe the telescope model adopted to mimic the response of the SKA-Low array; in Sec.\ref{sec:skymodel} we describe the model of the sky, including point-like and diffuse foregrounds and the CD/EoR signal; in Sec.\ref{sec:errormodel} we describe the error model, which includes corrupting effects introduced in the data. Sec.\ref{sec:products} describes the simulation outputs that we are making available; finally, conclusions are drawn in Sec.\ref{sec:conclu}. This paper is accompanied by data and of codes repositories, as specified in Sec. \ref{sec:repos}. 

\section{Telescope model} \label{sec:instrument}
We simulate full visibility data using the {\tt OSKAR}\footnote{\url{https://ska-telescope.gitlab.io/sim/oskar/}} \citep{oskar} package. {\tt OSKAR} makes use of a telescope model, containing the details of all antennas/stations, and a sky model, containing a list of all the sources in the sky. There is provision to include a variety of instrumental errors into the simulation, which we refer to as the error model. 

The parameters adopted for the {\tt OSKAR} simulation are as follows:
\begin{itemize}
\item field centre (RA,Dec) = (00h, -30deg);
\item 106--196\,MHz frequency range with frequency sampling of 100\,kHz. The chosen frequency range tracks the redshifted 21 cm line in the redshift range of $z=6.2$--12.4;
\item 4\,h track duration sampled with a 10\,s integration time;
\item thermal noise level consistent with an integration time of 1000\,h.
\end{itemize}

The choice of field is somewhat arbitrary and has not been optimised for an EoR observation. Sky components are mostly based on simulations, and therefore do not accurately represent the chosen area of the sky (see Sec.\ref{sec:skymodel} for more details).

\begin{figure}
\includegraphics[width=9cm]{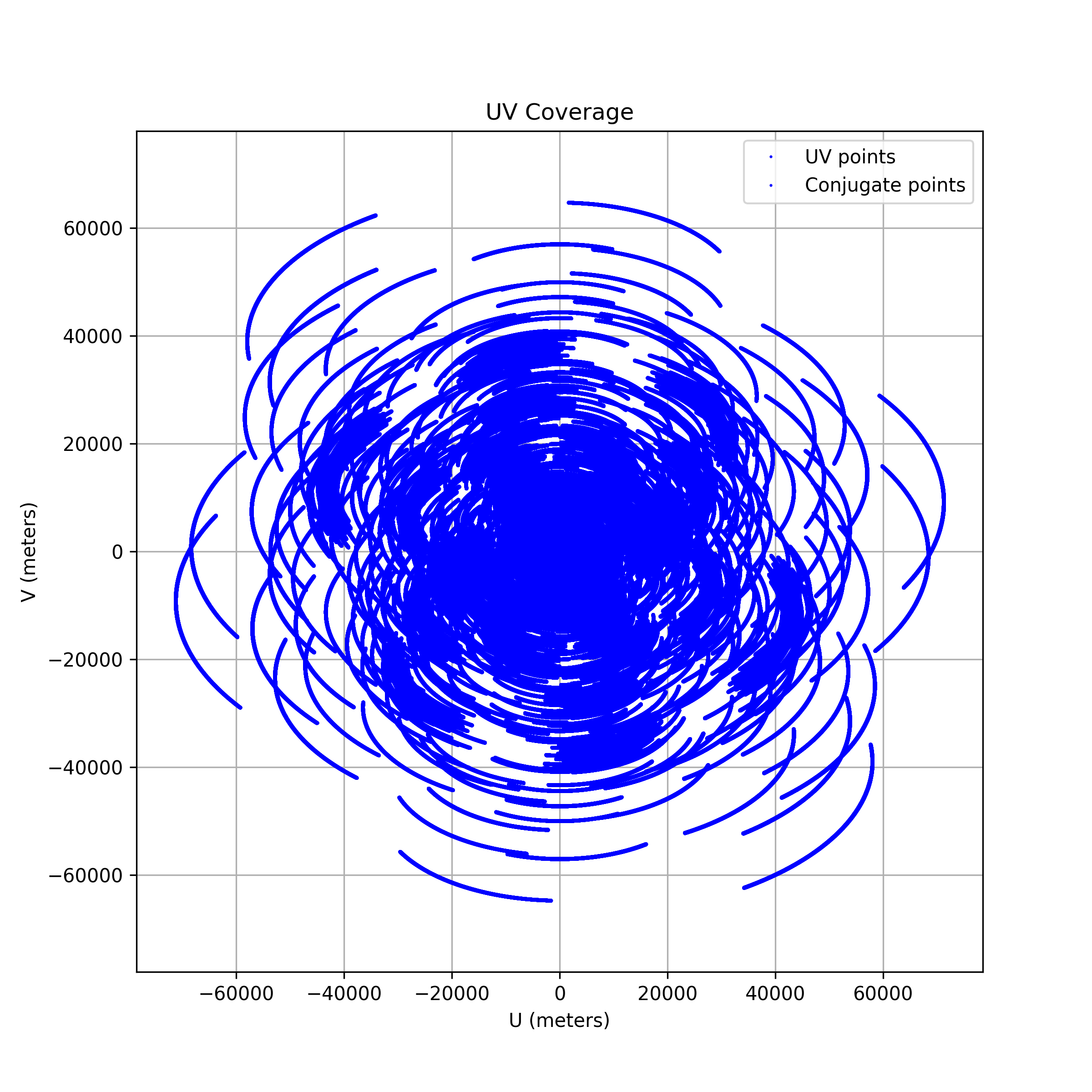}
\caption{$u,v$ coverage corresponding to the simulated observation. \label{fig:uvsampling}}
\end{figure}

The telescope model makes use of the SKA-Low configuration of 512 stations \citep{ska_low_conf}. The $u,v$ coverage  corresponding to this configuration is shown in Fig. \ref{fig:uvsampling}. This configuration has a maximum baseline of $B_{\rm max}\sim 74$ km which corresponds to a maximum spatial resolution of Full Width at Half Maximum (FWHM) $\sim 10$ arcsec in the frequency range explored here. 

The station layout is the so-called \emph{Vogel} layout, a one arm spiral configuration with a uniform areal density of the 256 dual-polarisation antennas and a maximally diverse azimuthal sampling \citep{vogel}. This is distinct from the "perturbed Vogel" layout that is actually being deployed within SKA-Low, in which a degree of randomisation has been added to the areal sampling. Only a single station layout has been adopted in our simulation to reduce computational expense, rather than a full set of 512 diverse station layouts. The implication is that the effective station beam for all visibilities is simply the auto-correlation beam of this one station layout. The most important consequence of this simplification is an elevated far side-lobe response. An attempt to compensate for this effect is made within the sky model definition (see Sec. \ref{sec:skymodel} for more details). The {\tt OSKAR} code already supports use of distinct station layouts throughout, so this could be adopted if a more realistic station beam response was desired for each simulated visibility.

The voltage response pattern of each antenna within each station has also been assumed to be the {\tt OSKAR} default, a half-wavelength (at each observing frequency) dipole pattern with a North-South/East-West orientation of the X/Y polarisations, in order to minimise computational expense. In reality, the embedded element voltage patterns within the SKA-Low stations are known to have significant variability in their shape that has a strong frequency dependence due to the varying degree of inter-antenna coupling between nearby antennas, while the intrinsic X/Y antenna polarisations have been given a distinctly different orientation within each of the 512 stations rather than being co-aligned with the cardinal directions. Nominal X/Y polarisation response will be determined during digital beam-forming for each station from this local X/Y starting point. Work is currently underway to embed realistic embedded element patterns for SKA-Low within the {\tt OSKAR} code, which would permit more realistic simulations to be undertaken.

\section{Sky Model}\label{sec:skymodel}
The sky components described in this Section are explicitly designed as a plausible realisation of the true sky rather than a prediction for the simulated field. This was done on purpose to ensure a blind foreground mitigation exercise. Some real data has been used, as noted below.

{\tt OSKAR} makes use of a sky model consisting of a list of sources modeled spatially as 2D Gaussians and spectrally as power laws. Although this might appear like a severe limitation, there are strategies that allow a much more general model of the sky to be used as an input. To overcome the power-law approximation for the frequency behaviour of the components, our sky model has been generated for each frequency channel, so that any frequency behaviour is accurately tracked. 

To introduce spatially diffuse components, we have produced gridded maps with an appropriate pixel size. Each pixel in the gridded representation has then been input as a Gaussian source model component in {\tt OSKAR}. We have used gridded representations for the EoR and the Galactic foregrounds, that are spatially diffuse, but also for the cumulative contribution from faint extragalactic sources. This has been done to reduce the computational cost of the simulation, which depends on the number of {\tt OSKAR} sky model sources. As a compromise between accuracy and computational complexity, we have adopted two grids: 

\begin{itemize}
\item \emph{Fine grid}: $5760 \times 5760$ pixels, which gives a $5 \times 5$\, arcsec pixel sampling over the FoV area. This cube was smoothed with a $10 \times 10$ arcsec FWHM gridding kernel and pixels from this grid were used in {\tt OSKAR} as Gaussian source model components of $10\times10$ arcsec extent with a flux density of the pixel brightness.
\item \emph{Coarse grid}: $512 \times 512$ pixels, which gives a $56.25 \times 56.25$\, arcsec pixel sampling over the FoV area. This cube was smoothed to an effective resolution of $112 \times 112$\,arcsec and pixels from this grid were used in {\tt OSKAR} as Gaussian source model components with spatial extent of $112 \times 112$\,arcsec and flux density of the pixel brightness. 
\end{itemize}

Below we list all the sky components and the way they have been modelled in {\tt OSKAR}.

\subsection{Extragalactic sources} \label{sec:exgal}
\subsubsection{Out-of-field sources}\label{sec:outer}
Sources that are bright enough for their contribution to be significant near the pointing direction even when entering the field through the telescope's far sidelobes were included over the full 2$\pi$ steradians above the horizon at any time. In this category are both the so-called "A-Team" sources that are brighter than a few 100\,Jy at 200\,MHz as well as all sources brighter than 5\,Jy at 150\,MHz (about 1200 in number) from the  \cite{2021PASA...38...57L} composite GLEAM and LoBES catalogue. 
These sources were all included individually as {\tt OSKAR} components, represented by an elliptical Gaussian approximation spatially and with a frequency-dependent flux density determined from an amplitude and spectral index defined at some reference frequency. 


\subsubsection{In-field sources}
In-field extragalactic sources have been included in numbers consistent with a sample complete down to 1\,$\mu$Jy  at 150\,MHz over a spatial extent of $8\times8$ degrees centred on the pointing direction. The strongest sources (above 100\,mJy at 150\,MHz) are from the \cite{2021PASA...38...57L} catalogue mentioned previously and were inserted as individual {\tt OSKAR} components. 

Sources with $1\,\mu\rm{Jy}\leq I_{150\,\rm{MHz}}<100\,mJy$ are from a simulated catalogue obtained with the T-RECS\footnote{\url{https://github.com/abonaldi/TRECS}} \citep{2023MNRAS.524..993B,2019MNRAS.482....2B} code. The T-RECS catalogue was converted to an image cube of $5 \times 5$\,arcsec pixel size with a dedicated code\footnote{\url{https://github.com/abonaldi/SKAO-SDC}} which represents resolved star-forming galaxies as exponential Sersic profiles, steep-spectrum AGN with a library of real AGN images, flat-spectrum AGN and unresolved sources with a Gaussian representation. Pixels in this gridded representation exceeding $10^{-5}$\,Jy/pixel (some 154,000 in number) were used in the fine grid; the remaining ones were spatially smoothed to an effective resolution of $112 \times 112$\,arcsec and included in the coarse grid. 

\begin{figure}

\includegraphics[width=9cm]{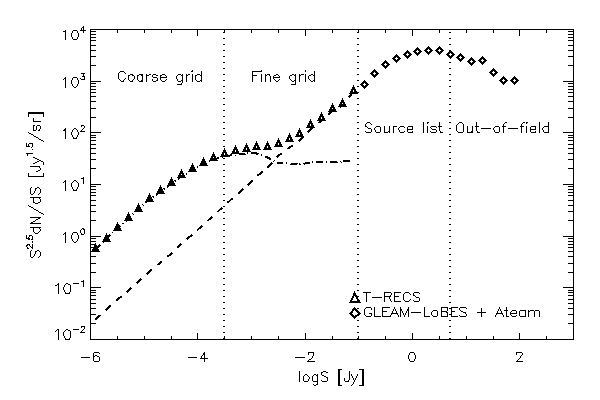}
\caption{Differential source counts for the extragalactic sources inserted in the simulation. We also indicated the flux limits used for the out-of-field sources, the sources included as a source list, and those included as a coarse and fine grid. For the T-RECS sources (triangles) the counts are also shown separately for AGN (dashed line) and SFGs (dash-dotted line). In total we included over 15 million sources. \label{fig:counts}}
\end{figure}

In figure \ref{fig:counts} we show the number counts of the sources at 150\,MHz, their provenance, and the way that they were modeled within {\tt OSKAR}. In total we have more than 15 million sources, which motivated the use of a gridded representation for the fainter sources, to reduce the computational complexity of the simulation. 

\subsection{Galactic diffuse emission}
Our Galactic diffuse total emission is based on the Galactic Sky Model 2016 \citep[GSM2016,][]{2017MNRAS.464.3486Z}, evaluated at each observing frequency towards the specified pointing centre. The PyGDSM implementation\footnote{\url{https://github.com/telegraphic/pygdsm}} \citep{2016ascl.soft03013P} of this sky model relies on a linear interpolation in log(frequency) between the reference frequencies of the model (specifically 85, 150, and 408\,MHz for our case). We modified this code to instead provide a quadratic interpolation in log(frequency) in order to eliminate discontinuous derivatives within our band. The GSM2016 model is limited in its effective angular resolution by the native resolution of the all-sky observations that it employs. At the low radio frequencies of relevance for this work, this is about 1\,degree. We have supplemented our model of the Galactic foreground emission by including total synchrotron emission at the relevant radio frequencies from a magneto-hydrodynamical (MHD) simulation \citep{2022A&A...663A..37B} of a hundreds-of-parsec wide Galactic volume sampled with 512 $\times$ 512 pixels. These synthetic data, derived from \citet{2017A&A...599A..94N}, represent a realistic model of a shock-driven turbulent, magnetized, and multiphase interstellar medium, whose dynamics are affected by stellar feedback. The spatial frequency content of the GSM2016 and MHD sky models was merged using the {\tt miriad}\footnote{\url{https://www.atnf.csiro.au/computing/software/miriad/}} \citep{1995ASPC...77..433S} package task {\tt immerge} with a normalisation at overlapping spatial scales, that was tied to the GSM2016 sky model at each frequency. In this way, filamentary structures in the Galactic foreground model were extended into the arcmin regime and the result was scaled to units of Jy/pixel. The Galactic diffuse emission was added to the coarse grid.

\subsection{Cosmic Dawn and Epoch of Reionisation}
\begin{figure}
\includegraphics[width=8.5cm]{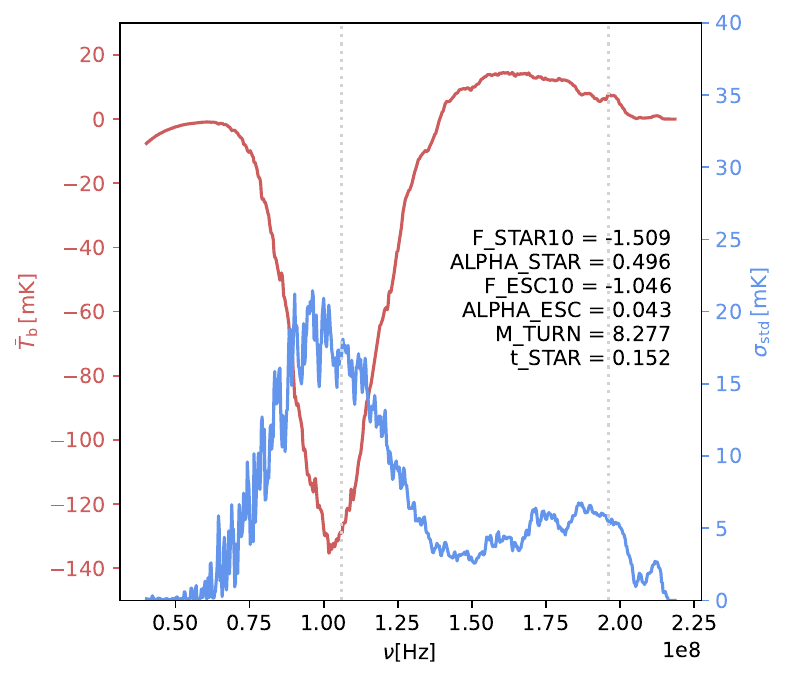}
\caption{{\tt 21cmFAST} parameters used for the generation of the EoR signal and corresponding mean (red) and standard deviation (blue) of the brightness temperature as a function of frequency. The vertical dotted lines represent the first and last frequency channel in the simulation.}
\label{fig:global}
\end{figure}

\begin{figure*}
\includegraphics[width=16cm]{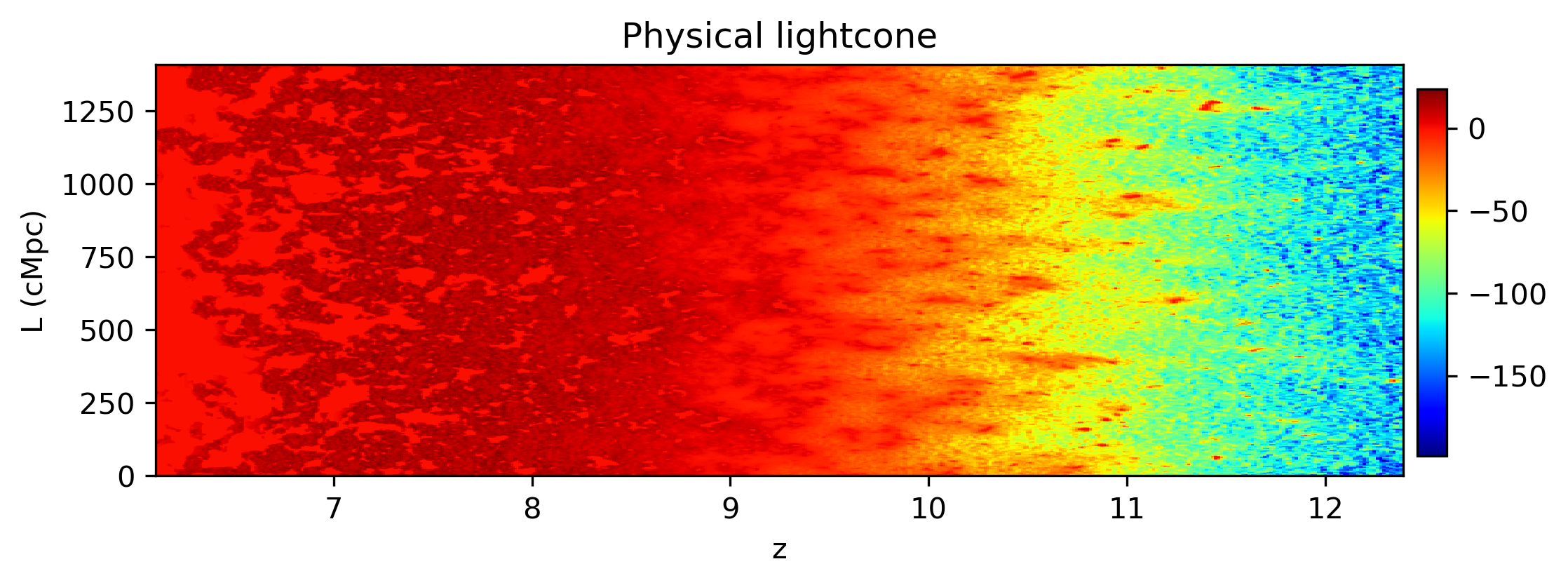}
\caption{Lightcone in physical quantities corresponding to the chosen reionisation scenario over the redshift range of the simulation. }
\label{fig:cone}
\end{figure*}

Finally, a {\tt 21cmFAST}\footnote{\url{https://21cmfast.readthedocs.io/en/latest/}} \citep{Murray2020,2011MNRAS.411..955M} simulation of the Cosmic Dawn and Epoch of Reionisation (CD/EoR) was generated. We have adopted {\tt 21cmFAST} version 3.2.0 and Planck18 cosmology $\Omega_m=0.30964$, $\Omega_\Lambda=0.69036$, $H_0=67.66$. The {\tt 21cmFAST} reionisation parameters and the corresponding Brightness temperature are shown in Figure \ref{fig:global}. This scenario has been chosen since it is both physically plausible (given current observational constraints) and yields a strong CD/EoR signal over a wide frequency range, therefore providing good potential for detection and measurement. The resulting physical lightcone is in Figure \ref{fig:cone}. At the highest redshifts (lowest frequencies) there is CD signal, where $x_{\rm HI}$ traces the matter density. At the lowest redshift (highest frequency) the signal is dominated by reionisation. The neutral fraction is $\sim 90$\%, 50\%, 10\% at $z=9.6$, 6.8 and 6.3 ($\nu \sim 134$, 181, 196\,MHz), respectively.  

The simulated EoR signal, in units of cubic Mpc, has been converted into observational quantities (deg, deg, MHz) by means of the {\tt tools21cm} library\footnote{\url{https://tools21cm.readthedocs.io/}} and added to the coarse grid.

\begin{figure}
\includegraphics[width=9cm]{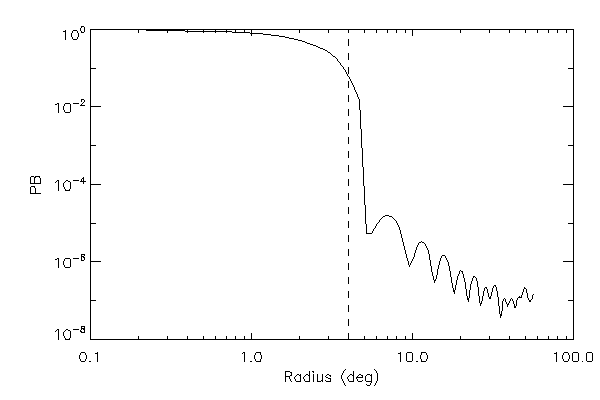}
\caption{Primary beam response at 106\,MHz, which shows the effective side-lobe level used to represent successful de-mixing. The vertical line shows the edge of the 8$\times$8 deg field.  \label{fig:pb}}
\end{figure}

\begin{figure*}
\includegraphics[width=16cm]{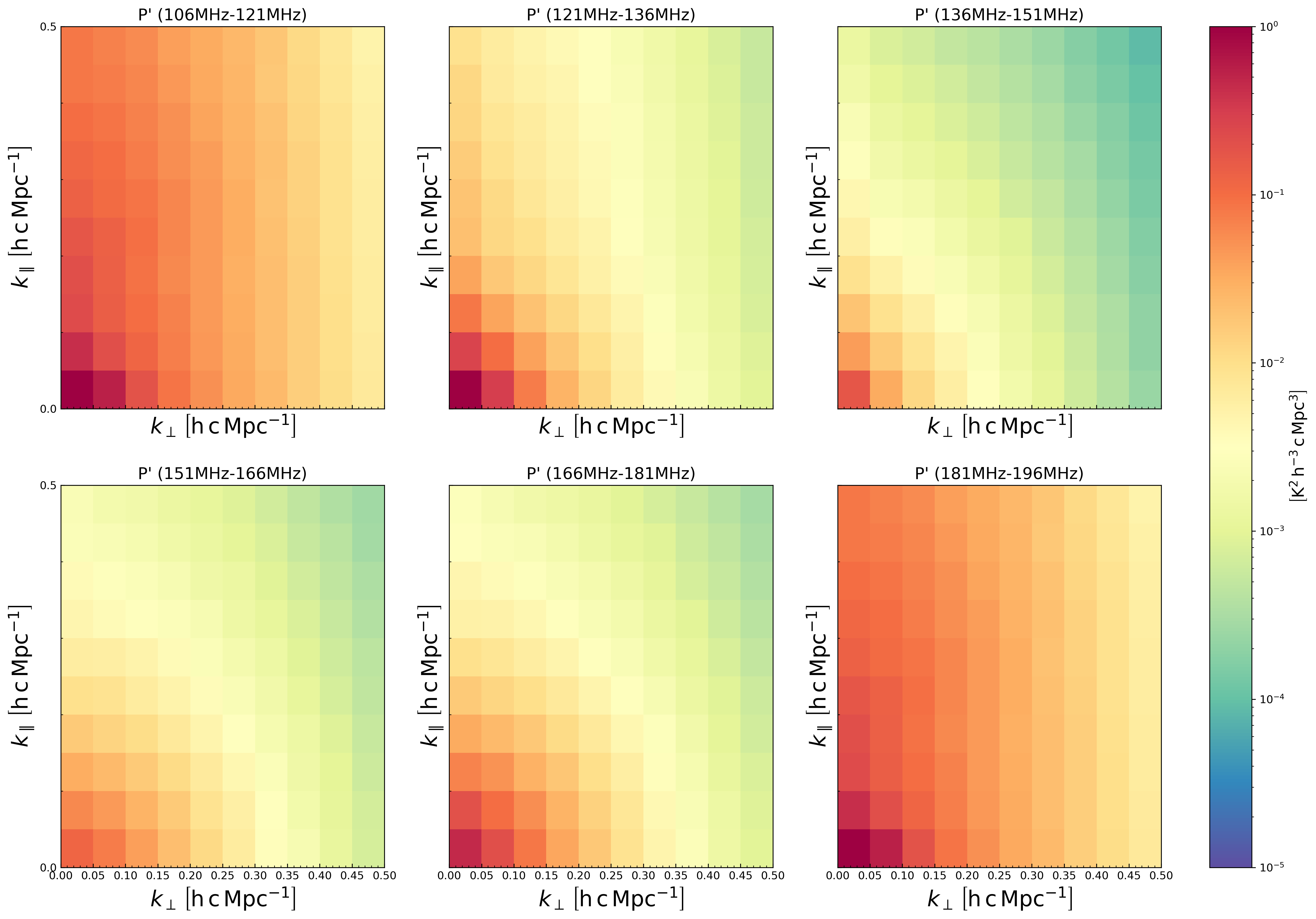}
\caption{Cylindrical power spectrum of the true noiseless EoR, $P'$.}
\label{fig:truth}
\end{figure*}

\section{Error Model}\label{sec:errormodel}
\subsection{Partial de-mixing of out-of-field sources} \label{sec:demix}
The out-of-field sources introduced in Sec. \ref{sec:outer} would normally be modelled and removed with a so-called “de-mixing” process within a calibration and imaging pipeline. Within our simulation these sources are attenuated by the {\tt OSKAR} station beam as defined Sec. \ref{sec:instrument}. However, as noted there, our use of identical rather than fully diverse station layouts yields significantly higher far-side-lobe levels, by about a factor of 10 on average, than would actually be the case. The visibility response to these sources will therefore be enhanced by this factor in the simulation. Our modelling of the de-mixing process (summarised in Table 3 of \citealt{SKA-TEL-SKO-0000941-02}) suggests that only the brightest 10 - 20 discrete sources above the local horizon will need to be included in the de-mixing process and we might realistically anticipate achieving a net reduction of their effects by a factor of $\sim 10^{-2}$. Since inclusion of an actual de-mixing pipeline was deemed beyond the scope of our current simulation efforts, we have instead attempted to approximate the net effect of both the presence of out-of-field sources as well as their partial de-mixing by defining an "effective" primary beam response, whereby the Sky Model is attenuated at large radii by a factor of $10^{-3}$ relative to the {\tt OSKAR} station beam to take account of both of these effects. The effective primary beam response is shown in Fig. \ref{fig:pb}. Although this far-field attenuation factor is acknowledged to be quite arbitrary in magnitude, we have explored its plausibility by running subsets of the simulation with both larger and smaller factors. Higher levels of attenuation completely eliminate the "pitch-fork" effect \citep{2015ApJ...807L..28T} seen in delay spectra due to residual out-of-field contamination, while lower levels of attenuation lead to much stronger contamination.

\subsection{Ionospheric effects and DD calibration errors} \label{sec:DDcal}
Another challenge that must be overcome in observations of this type is the successful modelling and elimination of propagation effects introduced into the visibility data by the ionosphere. We have used the {\tt ARatmospy}\footnote{\url{https://github.com/shrieks/ARatmospy}} \citep{2015OExpr..2333335S} code to construct an ionospheric model that was intended to represent moderately good observing conditions, characterised by a correlation scale, $r_0 = 7$\,km. 
The code is used to construct a time evolving phase screen above the telescope site that introduces Direction Dependent (DD) calibration errors into the visibilities via {\tt OSKAR}. Several ionospheric layers were simulated at plausible elevations, propagation speeds and propagation directions with a deliberately slow time evolution. It has been shown in Table 3 of \cite{SKA-TEL-SKO-0000941-02} that successful DD self-calibration should be possible between 100 and 200 MHz using the expected density of compact background sources (about 3 $\times 10^4$ in number exceeding 1.5 mJy of apparent flux density) with sufficient signal-to-noise per visibility within each four hour tracking observation. We can conservatively expect an improvement of the DD gain solutions by at least a factor of 10 in each observation. Since the total observing campaign being simulated consists of 250 repetitions of this basic four hour track and each should have uncorrelated ionospheric distortions, we can expect a further $\sqrt{250}$ improvement in the final DD calibration precision, over that obtained per track. We have attenuated the total electron counts of the single track observation and the corresponding phase modulation by the factor $10^{-2}$ as an estimate of this final outcome. 

\subsection{DI calibration errors} \label{sec:DIcal}
Another component of calibration pertains to the so-called Direction Independent (DI) gain calibration of the telescope stations as a function of both time and frequency. We have constructed a gain model for the telescope array which represents the amplitude and phase response of each of the 512 stations for every time interval and every frequency channel of the observation. We have fixed the mean amplitude response at unity and the mean phase response at zero, so that there are no systematic calibration errors, however an error distribution applies. The time and frequency fluctuations are assumed to be independent of each other and the gain error at each specific (time, frequency) is given by the complex product of the two, one-dimensional distributions. 

The one-dimensional gain distributions in each of time and frequency are first assigned random values from a Gaussian distribution with a specified standard deviation in each of amplitude and phase. We then make use of the {\tt colorednoise} python code\footnote{\url{https://github.com/felixpatzelt/colorednoise}} to take these “white-noise” distributions that have equal fluctuation power on all sampled scales and produce “red-noise” distributions with a $-2$ power-law index, whereby the greatest fluctuation power is present on the longest time intervals and over the largest frequency increments. Both the standard deviation and mean values are preserved in this process of colouring the noise. The resulting fluctuations are reminiscent of gain error patterns that are encountered in actual observations. 

In assigning a numerical value to the standard deviation of the DI amplitude and phase errors we have considered both what might be realistically achieved for the calibration quality of a single observation as well as a plausible degree of improvement that might be achieved by averaging the outcomes of a large number of independent observations. Specifically, we have simulated a continuous four-hour duration tracking observation. However, we are using this to simulate the outcome of a 1000 hour deep integration, which would be achieved in practice by carrying out 250 repetitions of such a four-hour track. The relevant parameters for self-calibration of the DI gains are also shown in Table 3 of \cite{SKA-TEL-SKO-0000941-02}. We can expect to achieve an RMS calibration precision of about \citep{1999ASPC..180..187C}, $\phi = \sigma/(\sqrt{N} S)$, where $\sigma$ is the visibility noise per solution interval (about 1 Jy over 100 kHz and 60 sec \citep{2019arXiv191212699B}), $N$ is the number of stations (512) and $S$ is the peak compact source brightness that occurs by chance within the field (about 20 Jy at 100 MHz). This yields an expected DI calibration precision of about 0.2 degrees of phase and 0.2\% in amplitude for an individual track. After averaging over the 250 tracks we assume residual errors of 0.02 degrees in phase and 0.02\% in amplitude for each of the time and the frequency domains. This is a reduction by a factor of 10 from the single track estimate, rather than the factor of 16 that would follow from $\sqrt{250}$, since we assume the solutions are not fully independent of one another, depending as they would on the same local sky model of the field. The time domain represents residual broadband gain calibration errors and the frequency domain represents residual bandpass calibration errors. While these estimates may prove to be optimistic, they at least provide a benchmark against which to assess the EoR signal recovery, and the simulation framework provides the means to test other circumstances. 

\subsection{Thermal noise}
Finally we have added a level of thermal noise to the visibilities that is meant to represent the anticipated sensitivity of the system in conjunction with the total assumed integration time. What is provided to {\tt OSKAR} is a file containing the SKA-Low station sensitivity expressed in units of Jy for a number of specific observing frequencies based on \citet{2019arXiv191212699B}. This nominal sensitivity per visibility, polarisation, unit bandwidth and unit time (about 1 Jy over 100 kHz and 60 sec) is then scaled for any possible differences between what is being simulated directly and what is being inferred. Specifically, only a single polarisation and four hour observation are being simulated directly, while the average of two polarisations and a total observing time of 1000 hours are being inferred. With this scaling in place, an appropriate thermal noise level is included in each visibility that is calculated.
\begin{figure*}
\begin{centering}
\begin{tabular}{cc}
\includegraphics[width=8.5cm]{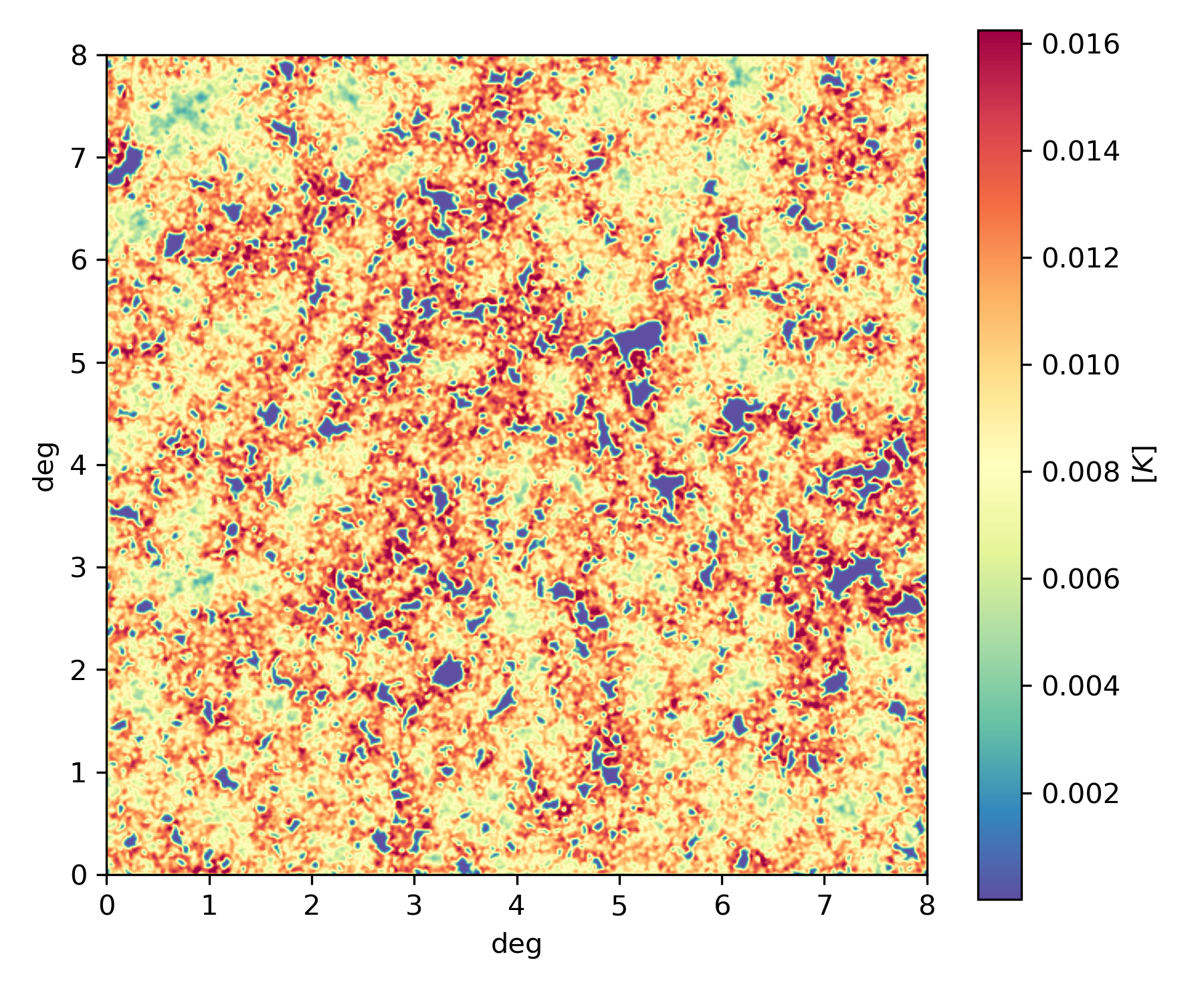}&\includegraphics[width=8.5cm]{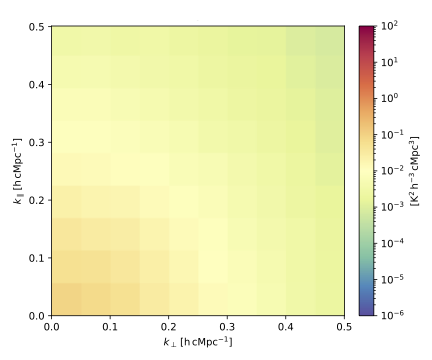}\\
\includegraphics[width=8.5cm]{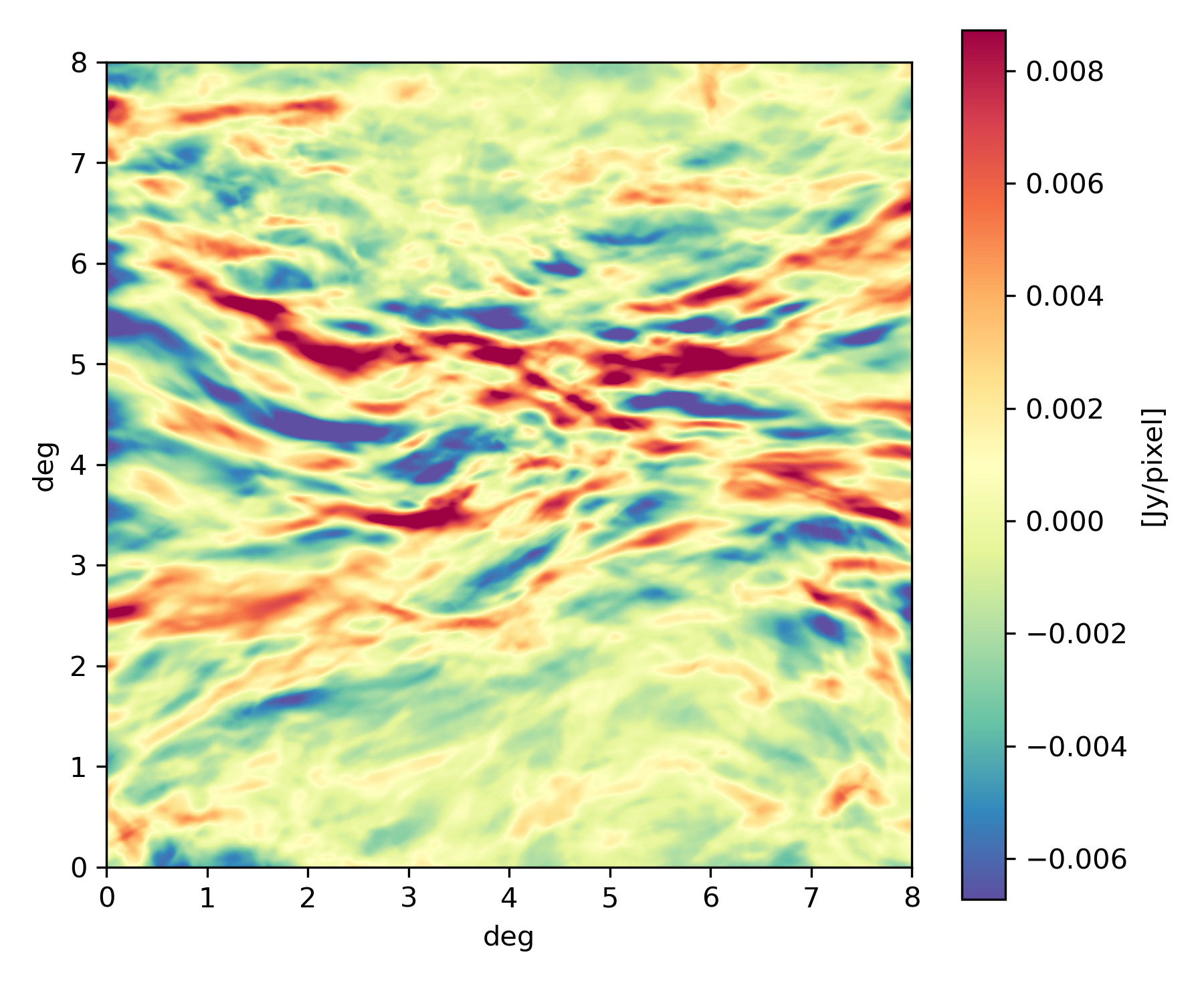}&\includegraphics[width=8.5cm]{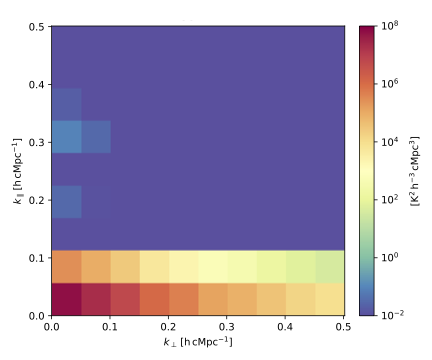}\\
\includegraphics[width=8.5cm]{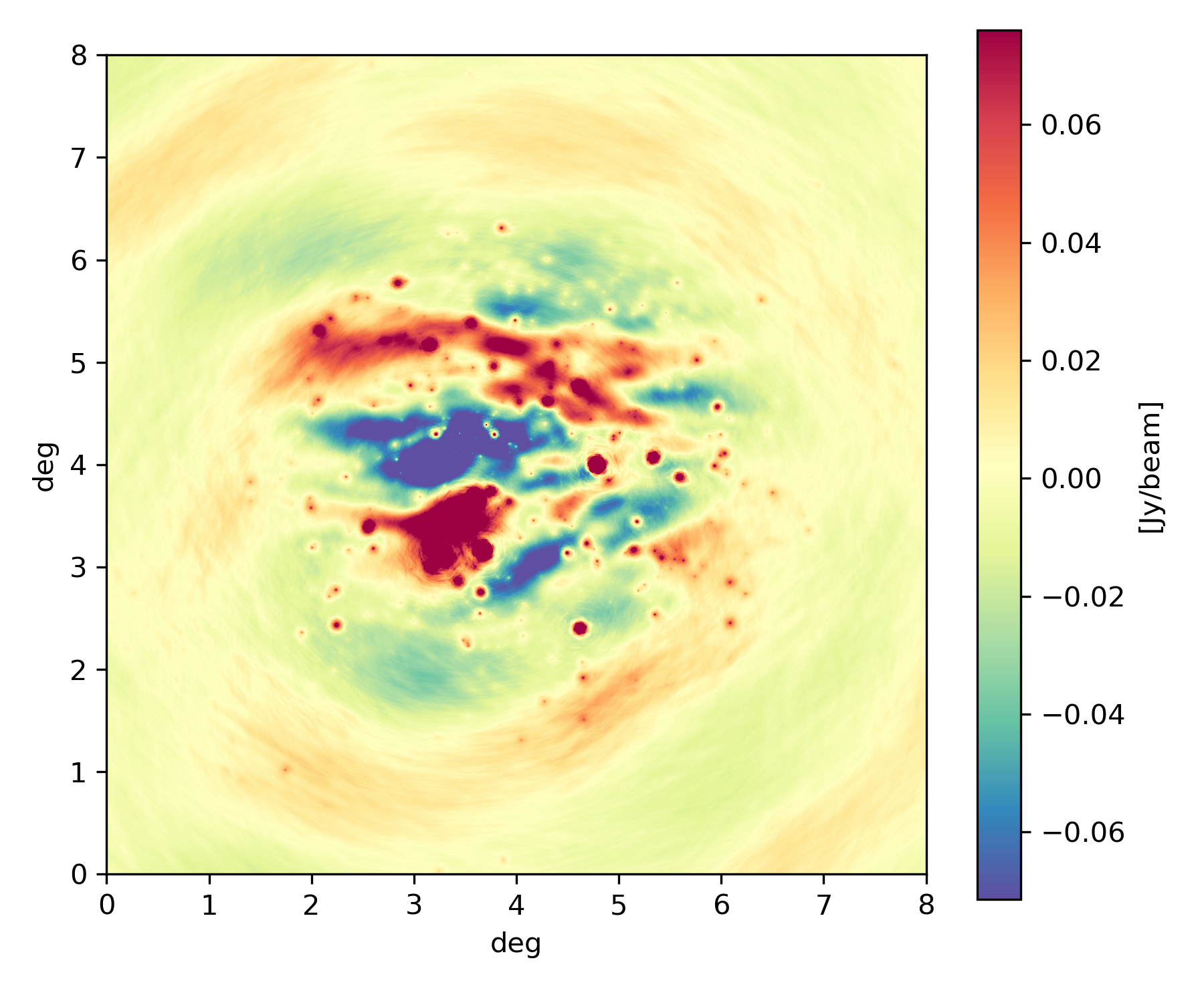}&\includegraphics[width=8.5cm]{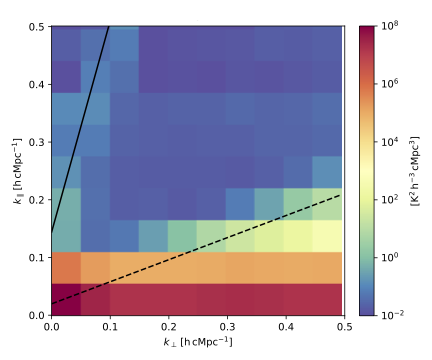}\\
\end{tabular}
\end{centering}
\caption{151\,MHz maps (left panels) and 2D power spectra at 151-166\,MHz (right panels) of: EoR component (top), Galactic emission component (middle); total contribution of all emission components and instrumental response} (bottom). Lines in the bottom-right panels represent the the horizon limit (solid black line) and the boundary corresponding to emission within the primary beam of the telescope (dashed black line).
\label{fig:comp}
\end{figure*}

\begin{figure}
\includegraphics[width=10cm]{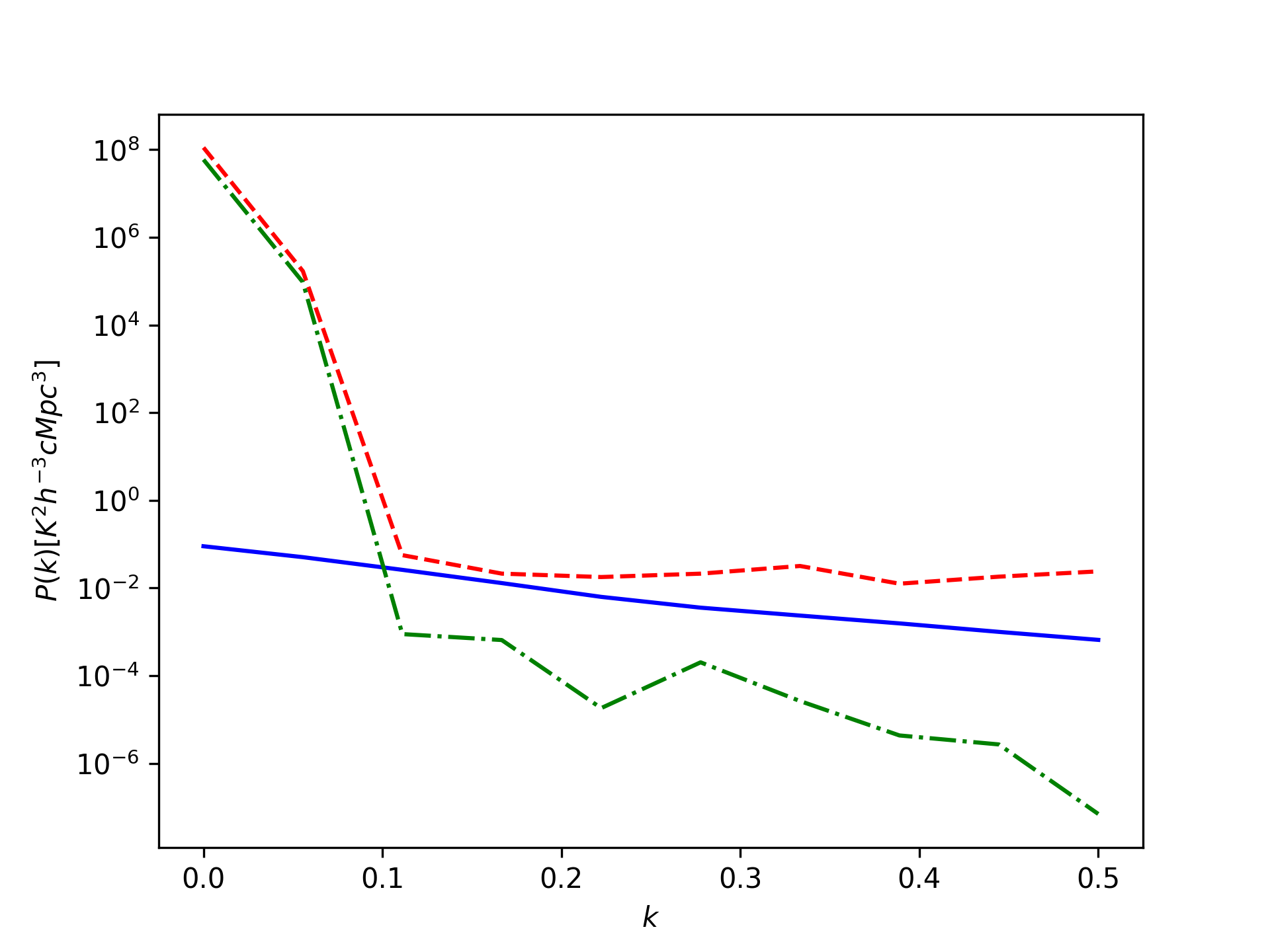}
\caption{ Diagonal terms ($k_\parallel=k_\bot$ elements only) of the 2D power spectra in the right panels of Figure \ref{fig:comp}: EoR component (solid line); Galactic emission component (dotted line); total contribution of all emission components and simulated instrumental response (dashed line). The EoR and the Galactic emission are presented as modelled and are free from simulated instrumental effects. The difference between the dotted line and the dashed line is therefore due to the combined effect of the EoR, the extragalactic sources, the ionosphere and the instrumental response.
}
\label{fig:diag}
\end{figure}

\subsection{Limitations} \label{sec:limits}
Although our EoR simulation provides an enhanced level of realism over previous work in this area, it is acknowledged to be far from complete. 
We have already drawn attention to several simplifications that were made to limit computational cost, including the use of only a single four-hour track to generate the visibility data, rather than the 250 repetitions of a track, each with distinct ionospheric conditions and its own DD and DI calibration challenges. Moreover, even this single track makes use of only a single station layout of antennas (rather than the 512 different variants that will be present) and only a single polarisation state of a highly idealised antenna element response. The deliberate diversity of both the station layouts and the nominal antenna polarisation states within the deployed SKA-Low is intended to regularise the instrumental response and reduce systematic errors. As noted in Sec. \ref{sec:instrument}, {\tt OSKAR} already provides support for the use of distinct station layouts and is anticipated to provide support for realistic embedded element patterns and non-cardinal alignment of the polarised response. It will be important to explore how these aspects of the deployed system flow through to the detectability of subtle astrophysical signals.
Another aspect of our simulation that will be important to extend is our very approximate treatment of out-of-field sources. It's clear that out-of-field sources can dominate the visibility response at these low frequencies. As described in Sec. \ref{sec:demix} we have modelled both the actual visibility response (from a distinct cross-correlation voltage beam for each visibility) as well as the process of successfully running a de-mixing pipeline by a simple attenuation of the auto-correlation station beam at large radii. This is acknowledged to be a very crude first step. The assumed far-field attenuation factor is at least available within the simulation code to allow further exploration of this phenomenon in the short term.

\section{Simulated products} \label{sec:products}
\subsection{Simulated Visibilities}
Visibilities are provided in the Measurement Set format, with every frequency channel being saved separately. The {\tt casa} package\footnote{\url{https://casa.nrao.edu/}} \citep{TheCASATeam_2022} was used to place the “SKA1-LOW” value into the “telescope” header with task {\tt vishead}. 
 The visibility data size amounts to 7.5\,TB.

\subsection{Simulated Images}
The {\tt wsclean} task\footnote{\url{https://gitlab.com/aroffringa/wsclean}} \citep{2014MNRAS.444..606O} was used to produce images and the corresponding synthesised beams (PSFs) from the visibilities, with natural weighting and no deconvolution. 
More specifically, the relevant {\tt wsclean} parameters were: -use-wgridder, -oversampling 4095, -kernel-size 15, -nwlayers 1000, -grid-mode kb, -taper-edge 100, and -padding 2. These values were chosen to account for the intrinsically 3D data acquisition to yield an accurately position-invariant PSF. Once images and PSFs for all frequency channels were available, these were assembled into a cube. The {\tt wsclean} specific fits header keywords are only preserved for the lowest frequency image that is assembled into each cube. 
The image cubes have a 2048 $\times$ 2048 pixels on a side which corresponds to a pixel size of 16 $\times$ 16 arcsec.

\subsection{EoR products}
Our list of products include the noiseless EoR signal, which can be used as a truth table to test foreground mitigation approaches. To this end we used the EoR cube in observational quantities derived at the end of Sec. \ref{sec:skymodel}. This has been smoothed to an effective resolution of $112 \times 112$ arcsec and sampled by 512 $\times$ 512 pixels, to reproduce the processing applied to all components in the coarse grid. 

A set of 6 cylindrical (2D) and 6 spherical (1D) power spectra were computed with the {\tt tools21cm} library on the noiseless EoR on 15\,MHz frequency bins.  
The 2D EoR power spectra for all considered frequencies in shown in Fig. \ref{fig:truth}.  

\section{Product inspection}
Figure \ref{fig:comp} shows the maps at 151\,MHz (left panels) and the corresponding 2D power spectra between 151 and 166\,MHz (right panels) for, from top to bottom, the clean EoR, the Galactic emission, and the total emission (EoR, Galactic and extragalactic emission, ionosphere). Instrumental effects (primary beam, PSF, calibration errors, etc.) are included in the total emission only. Colour scales for individual panels have been adjusted to allow inspecting each component, and are in general different. Note in particular that the clean EoR PS (top right) has a colour scale 6 orders of magnitude smaller than the Galactic and total PS components. To allow a comparison of the different amplitudes on the same scale, Fig. \ref{fig:diag} shows the diagonal terms ($k_\parallel=k_\bot$ elements only) of the 2D power spectra of Figure \ref{fig:comp}.

Figures \ref{fig:comp} and \ref{fig:diag} illustrate and quantify the cumulative effect of multiple foreground components and instrumental effects on the EoR signal. The Galactic component alone (middle panels in Fig. \ref{fig:comp} and dot-dashed line in Fig. \ref{fig:diag}) already dominates over the EoR signal at low $k$s, but leaves a relatively clean window for EoR detection over $k_\parallel\sim 0.1$ if instrumental effects are not considered. However, the addition of the extragalactic point sources introduces contamination at larger $k_\parallel$, and the ionosphere and instrumental effects cause contaminated modes to leak further into the cleaner part of the window (bottom panels of Fig. \ref{fig:comp} and dashed line in Fig. \ref{fig:diag}). 

While the current simulation does not fully capture the complexity of real data (see Sec. \ref{sec:limits} for a list of the main simplifying assumptions), it constitutes a significant step forward with respect to most simulation work in the literature. It is therefore a useful benchmark against which to test foreground cleaning and avoidance methods that are either being developed or applied to real data, as demonstrated by the SKA SDC3a data challenge \citep[][]{sdc3a2}.

\section{Conclusions} \label{sec:conclu}
We have presented a comprehensive and realistic SKA-Low simulation designed to be used to assess the efficacy of EoR foreground-mitigation techniques. Our sky model includes the CD/EoR component generated with the {\tt py21cmFAST} simulation package \citep{Murray2020}; Galactic foregrounds from the GSM2016 model  \citep{2017MNRAS.464.3486Z} and with small scales added above the native $\sim 1$\,deg resolution with the MHD simulation presented in \cite{2022A&A...663A..37B}; strong out-of-field extragalactic sources and in-field sources from the \cite{2021PASA...38...57L} catalogue; faint extragalactic sources (down to 1\,$\mu$Jy at 150\,MHz) from the T-RECS \citep{2019MNRAS.482....2B,2023MNRAS.524..993B} simulation. 

The telescope simulation was fully performed in the visibility domain with the {\tt OSKAR} \citep{oskar} code. The telescope configuration corresponds to the full SKA-Low array (512 stations, \citealt{ska_low_conf}). A 4\,h track  was simulated, and the noise finally scaled to a 1000\,h total integration. 

Our simulation includes partial de-mixing of out-of-field strong sources; ionospheric effects generated with the {\tt ARatmospy} \citep{2015OExpr..2333335S} software; DD and DI calibration errors. 
Imaging involved {\tt wsclean} and further processing into power spectra used the {\tt ps\_eor} and {\tt tools21cm} libraries. 

The interaction between instrumental effects and the sky signal is known to complicate the structure of the components both in the frequency and spatial domains, partially violating assumptions made by foreground-removal techniques. Those effects therefore need to be included to be able to test foreground mitigation in a realistic way. 
This simulation represents a step forward in this respect compared to previous work. As the basis of the SKA's foreground mitigation data challenge, SDC3a \citep[see][for a full description of the analysis and results]{sdc3a2} it has already yielded valuable lessons for the EoR data analysis community.

One of the outcomes of SDC3a is that multiple stages of analysis are needed to effectively decouple the different sources of complexity -- point-source and diffuse foreground removal, and disentangling instrumental effects from sky signals. An iterative detection and subtraction of strong point sources before diffuse foreground removal, or an explicit deconvolution/convolution step to restore the smoothness of the components in the frequency dimension, achieved overall the best results in the challenge. 
Another outcome of the challenge is that error estimation is generally not robust enough at present against residual systematics after foreground mitigation. 

The dataset released in this work, and the simulation framework published as a set of codes, can help further this research, as the community continues to pursue the detection of the EoR signal with currently operating and upcoming instrument, including the SKA.

\section{Data availability}\label{sec:repos}
The repositories used in this pipeline, including the scripts to reproduce this simulation, are listed in Table \ref{tab:repos}. Main dependencies include the {\tt tools21cm}, {\tt py21cmFAST}, {\tt ps\_eor} and {\tt OSKAR} codes, as well as common astronomical packages.

The data cubes are available at \url{https://www.skao.int/en/science-users/170/simulated-data-products/ska-low-eor}.

\begin{table*}
\caption{Main repositories for our simulation pipeline. \label{tab:repos}}
\begin{center}

\begin{tabular}{llll}
\hline
Code name&How is used in this pipeline&Repository Link\\
\hline
SDC3a\_repro&Main pipeline script and input files&\url{https://github.com/abonaldi/SDC3a_repro}&\\
T-RECS&Generate catalogue of extragalactic sources&\url{https://github.com/abonaldi/TRECS}\\
SKAO-SDC&Create an image cube from the T-RECS catalogue&\url{https://github.com/abonaldi/SKAO-SDC}\\
SKA Low simulator&Consolidate all the sky components with {\tt OSKAR}&\url{https://github.com/omkarbait/ska-low-simulator}&\\
\hline
\end{tabular}
\end{center}
\end{table*}

\section{Acknowledgements}
AB and EL thank A. Mesinger, A. Nasirudin, S. Giri, B. Greig, G. Mellema, Y. Qin for useful advice and support in setting up the {\tt 21cmFAST} simulation. AB thanks M. Bianco and S. Giri for help and support with the {\tt tools21cm} power spectrum estimation routines. FGM acknowledges support from the I-DAWN project, funded by the DIM-ORIGINS programme. ABracco acknowledges financial support from the INAF initiative "IAF Astronomy Fellowships in Italy" (grant name MEGASKAT). ABracco thanks V. Jeli\'c, E. Ntormousi, and M. Padovani for their contribution to the diffuse Galactic foreground simulation.  
\bibliography{sdc3b}

\newpage
\bibliographystyle{mnras}

\end{document}